\documentclass{elsart}
\usepackage{harvard}
\usepackage{graphicx}
\usepackage{amssymb}

\def\url#1{{\ttfamily\def\/{/\discretionary{}{}{}}#1}}

\begin{document}

\begin{frontmatter}

\title{The Gaseous Environments of Radio Galaxies}
\author{D.M. Worrall\thanksref{dmw}}

\address{Harvard-Smithsonian Center for Astrophysics, 
60 Garden Street, Cambridge, MA 02138, U.S.A.\\
\&\\
Department of Physics, University of Bristol, Tyndall Avenue,
Bristol~BS8~1TL, U.K.}

\thanks[dmw]{E-mail: d.worrall@bristol.ac.uk}

\begin{abstract}

X-ray emission traces the gaseous environments of radio sources.  The
medium must be present for jet confinement, but what are its influence
on jet fuelling, dynamics, propagation, and disruption?  The
observational situation is both complicated and enriched by
radio sources being multi-component X-ray emitters, with several
possible regions of non-thermal emission.  Recent work, primarily
based on sensitive ROSAT pointings, is used to contrast the X-ray
emission and environments of radio sources with (a) low power, (b)
high power at high redshift, (c) high power at lower redshift, and (d)
GHz peaked spectrum emission.  The trends in external gas density and
pressure near extended radio structures are reviewed.
Imminently-available X-ray measurements with vastly improved
resolution and sensitivity have great potential for resolving many open
issues.

\end{abstract}


\end{frontmatter}

\section{Introduction}
\label{intro}

A gaseous medium is essential for the propagation of a radio jet.
Although hot gas is traced through its X-ray thermal bremsstrahlung
and line radiation, the X-ray detection of gaseous environments around
most radio sources has required the sensitivity and angular resolution
available only in the last decade with ROSAT.  While it is clear that
an X-ray emitting medium must be present at some level for the well
being of the radio source, major concerns from a radio-astronomy and
jet-theory perspective are the extent to which the medium influences
radio-jet fuelling, dynamics, propagation, and disruption.  From an
X-ray astronomy perspective, it is of interest to determine whether or
not the X-ray environments of radio sources are special, and to use
radio sources to indicate the presence of X-rays associated with
large-scale structure.

The X-ray-emitting medium of radio sources is complicated to study
because radio sources are multi-component X-ray emitters and include
several possible regions of non-thermal emission.  These non-thermal
components are themselves an important probe of the physical
conditions of the central engine and radio beams.

Since careful separation of the various X-ray components is essential
before progress can be made, this paper first reviews non-gaseous
emission components in radio galaxies, and presents examples of
observational biases.  The discussion of the X-ray emitting
environments is broken down by radio-source power and morphology, and
trends in external gas density and pressure near the extended radio
structures are reviewed.  There are many open issues, and I will
identify the potential for resolving them using the new X-ray
missions with their vastly improved resolution and sensitivity.

Because the central X-ray regions of radio galaxies experience
anisotropic emission and absorption, assumptions concerning
Unification color an approach to the subject and can be tested by the
data.  This paper makes the standard Unification assumptions that
BL~Lac objects, radio-loud quasars, and broad-line radio galaxies
(BLRGs) are the unobscured, favorably-beamed counterparts of
low-power FRI \cite{Fr74} radio galaxies, high-redshift FRII
galaxies, and low-redshift FRIIs, respectively. $H_o = 50$ km s$^{-1}$
Mpc$^{-1}$ and $q_o = 0$ are adopted throughout.

\section{Multi-component X-ray Emission}
\label{multi}

\subsection{Cygnus A}
\label{cygnusa}

\begin{figure}
 \begin{center}
  \includegraphics*[width=7.5cm]{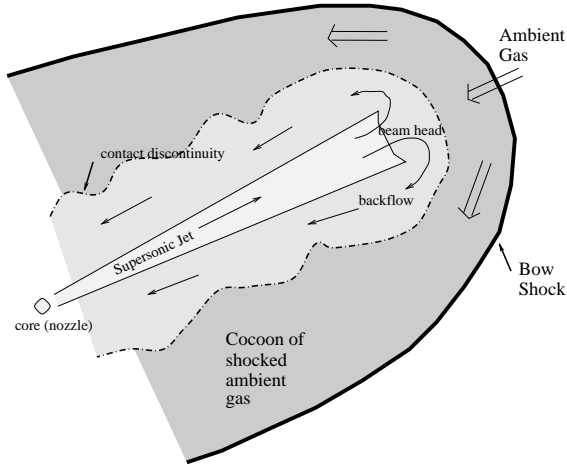}
 \end{center}
 \caption{Sketch of the termination region of a powerful radio jet
  viewed in the rest frame of the bow shock.  Radio lobe emission
  fills the region inside the contact discontinuity.  Between the
  contact discontinuity and the bow shock we expect the ambient
  X-ray-emitting medium to be both compressed and heated with respect
  to the medium in front of the bow shock.}
 \label{fig-beam}
\end{figure}

Cygnus~A is sufficiently powerful and close to show a full array of
X-ray emission components.  Its surrounding cluster emission is
strong, but when modelled and subtracted from the $\sim 0.2 - 2$~keV
ROSAT High Resolution Imager (HRI) image, \citeasnoun{Carill94} found
pronounced soft-excess emission associated with the radio hotspots,
core, and two (possibly three) regions around the limb of the lobe
plasma, together with an X-ray deficit in the inner lobes.  With
reference to a standard model for a powerful radio source with
supersonic jet (Fig.~\ref{fig-beam}), it is interesting to speculate
that the excesses around the lobes are parts of ring structures
pinching the contact discontinuity. There was no evidence for
increased X-ray emission due to higher gas density ahead of the beam,
in the cocoon between the contact discontinuity (containing the radio
lobes) and bow shock, as might confirm the standard model, but
\citeasnoun{Carill94} argue that the increased luminosity due to
higher density may be offset by the effect of heating, which would
tend to remove X-ray emission to an energy band above that to which
ROSAT is sensitive.

\begin{figure}
 \begin{center}
  \includegraphics*[width=6.8cm]{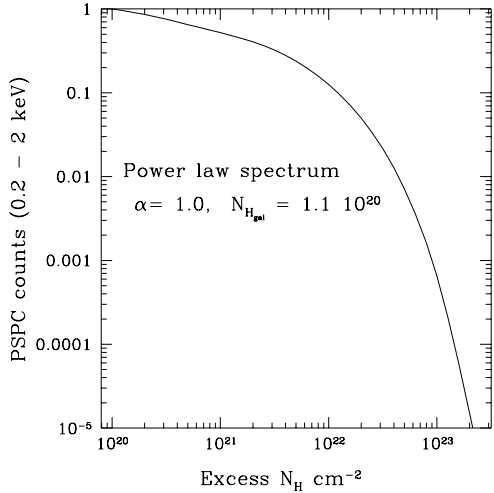}
 \end{center}
 \caption{As the excess (intrinsic) column density rises to more than
  a few $10^{22}$ cm$^{-2}$, the counts measured with ROSAT quickly
  fall.  The example is for the PSPC detector, assuming a source with
  a power-law spectrum of $\alpha = 1.0$ ($f_\nu \propto
  \nu^{-\alpha}$) and galactic column density of $1.1 \times 10^{20}$
  cm$^{-2}$, but a similar situation applies to the HRI and its
  measurement of the X-ray core of Cygnus~A. } 
 \label{fig-abs}
\end{figure}

\begin{figure}
 \begin{center}
  \includegraphics*[width=7.5cm]{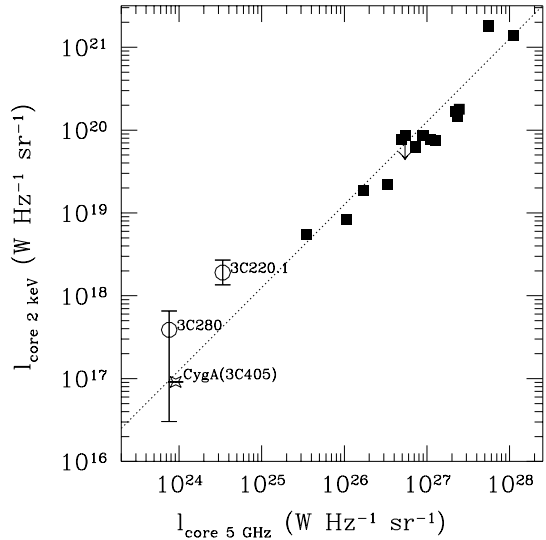}
 \end{center}
 \caption{The two high-redshift ($z > 0.6$) radio galaxies for which
core soft X-ray emission is separated from cluster emission, 3C~280
(Worrall et al.~1994) and 3C~220.1 (Hardcastle et al.~1998b), are
roughly consistent with an extrapolation of the core radio/X-ray
correlation for core-dominated quasars of comparable redshift (Worrall
et al.~1994).  Since these quasars are believed through Unification
models to be powerful radio galaxies oriented with their jets in the
line of sight (e.g.~Barthel 1989), the correlation supports the
interpretation of the core soft X-ray emission from these radio
galaxies as being beamed and associated with the radio jet.  Cygnus~A,
although local, has comparable radio-core luminosity to 3C~280, and
fits remarkably well on the correlation when the HRI core X-ray
emission is interpreted as radio-related.}
 \label{fig-cygcore}
\end{figure}

X-ray spectroscopy in the 2-10~keV energy band finds a poor fit to
cluster gas alone, and argues for the presence of non-thermal emission
seen through a large absorbing column, $N_{\rm H} \sim 4 \times
10^{23}$ cm$^{-2}$, and interpreted as emission from a heavily
obscured central AGN \cite{Arnaud87,Ueno94}.  Interestingly, this
absorbed core emission cannot be the soft-X-ray core excess in the
ROSAT HRI image \cite{Harris94b}, because such a high column density
has a disastrous effect on soft X-rays (Fig.~\ref{fig-abs}).  Instead,
the soft X-rays may arise from a central region in the radio source
where the only line-of-sight absorption is the Galactic column,
$N_{\rm H} \sim 3 \times 10^{21}$ cm$^{-2}$.  Indirect support for
this suggestion comes from the fact that the ratio of the unabsorbed
X-ray to core-radio luminosity is then very similar to that of
core-dominated quasars and those high-redshift counterparts of
Cygnus~A for which the core soft X-ray emission is separated from
cluster emission (Fig.~\ref{fig-cygcore}), although the cluster-scale
cooling flow in Cygnus~A \cite{Reyn96} should contribute at some level
to the HRI soft X-ray core excess.

\begin{figure}
 \begin{center}
  \includegraphics*[width=12.0cm]{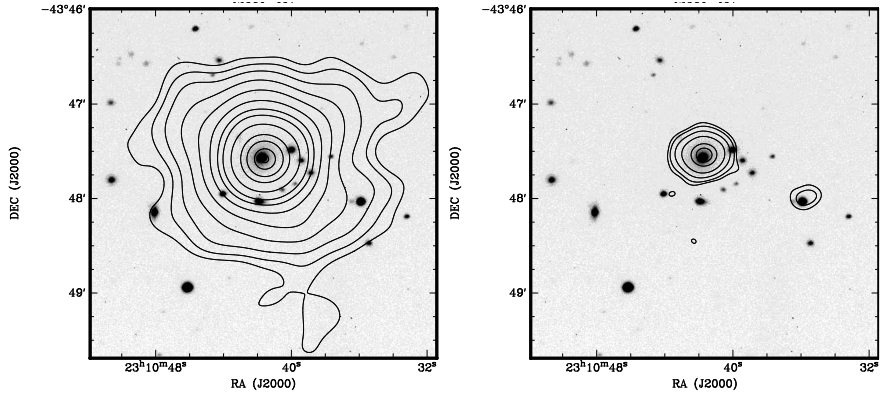}
 \end{center}
 \caption{The ROSAT PSPC and HRI are sensitive to different structures
in a complex source, such as J2310-437 (Worrall et al.~1999), a BL~Lac
object in an X-ray cluster.  The PSPC is more sensitive to low surface
brightness extended structures (left panel shows X-ray contours from a
4~ks PSPC exposure), whilst the superior spatial resolution of the HRI
pin-points features which are unresolved or of small spatial scale
(right panel shows 31~ks HRI exposure).  Lowest contours are at
$3.8\sigma$, and grey-scale is R-band CCD image.}
 \label{fig-2310}
\end{figure}

\subsection{Observational Biases}
\label{biases}

Because radio galaxies are multi-component X-ray emitters, the
energy-band, sensitivity, and spatial and spectral resolution of the
observing instrument influence what is measured.  Focussing X-ray
optics have the major advantage of decreasing the background, and so
{\it Einstein\/} was the first mission to detect some tens of radio
galaxies in soft X-rays \citeaffixed{Fab84}{e.g.} and to separate
components in nearby objects such as Cen~A and M~87
\cite{Feig81,Schr82}.  The mission contributing most to the subject
over the last decade has been ROSAT.  The combination of improved
sensitivity (roughly twice {\it Einstein's\/} collecting area at $\sim
1$~keV) and a longer mission have permitted relatively large samples
of radio galaxies to be studied, and the point response function (PRF)
of $\sim 15''$ Half Energy Width (HEW) for the Position Sensitive
Proportional Counter (PSPC) and $\sim 4''$ HEW for the HRI (although
with poorer sensitivity and no spectral resolution) has led to
component separation in many sources for which pointed observations
were made.  An object can look rather different when viewed with the
PSPC and HRI, with the former emphasizing extended emission and the
latter the compact components (Fig~\ref{fig-2310}).  ASCA has
increased the number of radio galaxies with {\it spectroscopic\/}
component separation \citeaffixed{Sambr99}{e.g.}  although, with its
relatively poor spatial resolution (HEW $\approx 3'$), weak sources
often give ambiguous results, with different combinations of spectral
models giving similarly good fits to the data.  The payload of
BeppoSax covers the broad energy range from soft X-rays to 300~keV,
although mostly with non-focussing optics \cite{Boell97}; its
strengths are therefore in broad-band studies of bright beamed
counterparts of radio galaxies, although there are tentative claims
for the detection of heavily-obscured AGN nuclei (as for Cygnus~A) in
some FRIs \cite{Truss98}.

Observational biases occur from redshift effects, not only in the
sense that in flux-limited samples the more distant sources are the
more powerful.  Extended X-ray emission tends to be seen around
high-redshift radio sources only if it is of cluster size and
strength; around low-redshift sources it is easier to detect the more
compact gaseous components than larger-scale emission which fills the
detector field of view.  Spectral measurements attempt to
measure excess absorption over that in the line of sight in our
Galaxy, and small fitted excesses become large intrinsic excesses when
transformed to the rest frame of a high-redshift source, where this
would not happen for a more local source.

\section{Non-gaseous X-ray Emission}
\label{nongaseous}

\subsection{Central Engine}
\label{engine}

The strength of central X-ray emission should depend on the accretion
process and black-hole mass, coupled with the effects of geometry and
absorption.  X-ray imaging has insufficient angular resolution to
separate such emission from beamed radiation associated with
an inner radio jet.  However, in sources where an absorbed power-law
component is present, it is easily recognized if it is dominant,
and a heavy excess absorption helps to make spectral separation
possible even if much of the emission is from
surrounding hot gas.  The best case is the hard X-ray detection of
emission from the core of
Cygnus~A (\S \ref{cygnusa}), and absorbed power-law components are
claimed in other radio galaxies with ROSAT, BeppoSAX and ASCA
\citeaffixed{Allfab92,Truss98,Sambr99}{e.g.}.  

Where the excess absorption is only modest, and this covers many of
the cases where the absorbed X-ray component is dominant, an
association with the central engine is questionable, and the absorbed
X-ray emission is most likely beamed emission associated with the
radio jet (\S \ref{beamed}).  This is illustrated by NGC~6251, where
the X-ray absorption of $\sim 10^{21}$ cm$^{-2}$ agrees both with that
inferred from reddening through the large-scale disk measured with HST
\cite{Fford99} and with an HI radio absorption-line measurement
\cite{Worrb99b}, and where the strength of X-ray relative to radio
emission in comparison with other radio galaxies argues independently
for a radio-related origin for the power-law X-ray emission
\cite{Worrb94}.  The `puzzling' excess absorption seen in BLRGs
\cite{Sambr99} might also be explained, at least in part, by cool gas
on larger scales than an inner torus absorbing jet-related
X-rays. This is consistent with the relatively strong X-ray emission
of BLRGs and the required orientation of their jets under Unification
models.

\subsection{Beamed X-ray Emission}
\label{beamed}

\begin{figure}
 \begin{center}
  \includegraphics*[width=7.5cm]{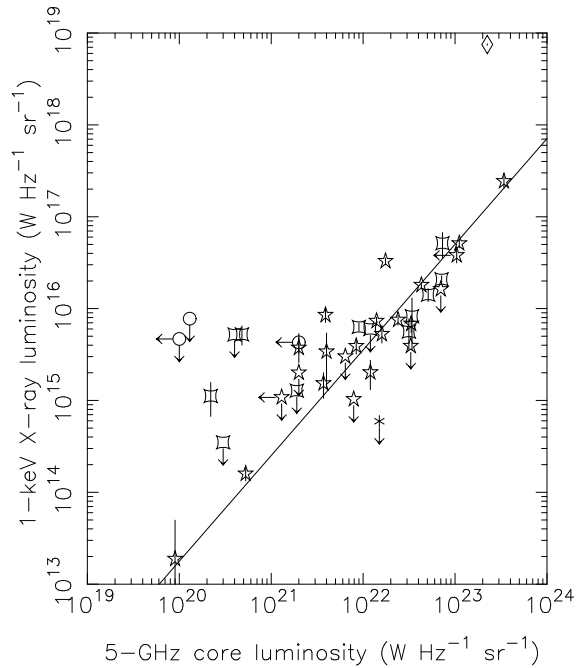}
 \end{center}
 \caption{Core X-ray {\it vs.} core radio luminosity for the sample of
B2 low-power radio galaxies observed with ROSAT in pointed
observations, after separation of any extended X-ray emission.  The
best-fit correlation taking into account non-detections (solid line)
excludes (based on astrophysical arguments) a starburst galaxy (cross)
and the broad-line radio galaxy 3C~382 (diamond) which fall in the
sample.  Radio galaxies in previously known optical clusters are shown
as squashed squares, relic radio sources are open circles,
and other sample members are shown as stars. Figure from Canosa et
al.~(1999). }
\label{fig-b2ll}
\end{figure}

ROSAT pointed observations have shown that the central soft
X-radiation of low-power radio galaxies is almost certainly dominated
by nonthermal emission associated with the radio jet.
\citeasnoun{Canos99} find that the core X-ray and radio emission are
well correlated in the B2 radio-galaxy sample (Fig~\ref{fig-b2ll}),
and a similar situation holds for the low-power 3CRR radio galaxies
\cite{Hardworr-3c99}.  M~87 and Cen~A are sufficiently close that
jet-related X-rays are resolved, and the fact that their X-ray to
radio ratio is similar to that for more distant unresolved X-ray cores
is further support for a jet-related origin of the core soft X-ray
emission in all such sources \cite{Worr97}.  Although in principle
such X-ray emission could be either synchrotron or inverse Compton in
origin, the relative proportions of radio, optical (HST) and X-ray
core emission, as compared with radio-selected BL Lac objects, argue
in favor of inverse Compton emission and predict a relatively flat
spectral index \cite{Hardworr-oc99}.  Flat-spectrum components
superimposed on thermal X-rays from hot gas are reported in the ASCA
spectra of several low-power radio galaxies, but are variously
interpreted as thermal emission associated with an advection dominated
accretion flow \cite{Allen99} and as higher than previously suggested
\citeaffixed{Fab89}{e.g.} X-ray emission from stellar and post-stellar
X-ray sources \cite{Mats97}.  Jet-related X-ray emission is also
likely to be a major contributor to the compact soft X-ray emission of
powerful radio galaxies \citeaffixed{Hardworr-3c99}{see
Fig~\ref{fig-cygcore} and}, although in general their greater distance
with respect to low-power sources leads to the expectation of a larger
contribution from extended gaseous emission within an unresolved X-ray
core.

\subsection{Non-thermal emission components away from the core}
\label{knots}

X-ray emission from compact radio hotspots has been detected in a
handful of sources, as summarized by \citeasnoun{Hard-hot98} and
\citeasnoun{Harr98}.  Such measurements are potentially of great
physical interest since the radio-emitting regions are normally well
localized and, if it can be shown that the X-rays are of inverse
Compton origin \citeaffixed{Harris94a}{e.g.}, the radio and X-ray
emission together probe the magnetic field strength and the balance
between particle and magnetic energy density.  A similar probe on
larger scales is provided through inverse Compton scattering of cosmic
microwave background (CMB) photons on particles in the radio lobes, with
detections reported in a few sources \cite{Feigel95,Tsak96,Tash98}.
\citeasnoun{Brun99}, in discussing extended emission in 3C~219, have
emphasized the role of AGN photons in Compton up-scattering, but at
present sources with extended X-ray emission of gaseous origin
vastly outnumber those for which extended inverse Compton X-ray
emission is likely to have been detected.

\section{The gaseous X-ray-emitting medium}
\label{gaseous}

\subsection{Questions and Timescales}
\label{timescales}

There are a number of interesting questions we would like to answer
when studying the gaseous X-ray emitting environment around a radio
source.  Is the external gas pressure greater than or equal to the
minimum pressure calculated for the jet, in which case alternative
methods of jet confinement are not required?  We would like to know if
the radio galaxy is moving in the X-ray medium (so that the jet is
affected by ram-pressure forces), or if there are large-scale gas
motions (cooling-flows, mergers, winds, etc) which affect the
production of jets or cause their disruption.  Is the gas distribution
smooth between large and small scales, or do abrupt transitions in
temperature and density induce observable radio deformations?  Do we
see evidence of direct interaction between the jets and the
surrounding medium (e.g., heating of the X-ray gas) which can be
tested against model predictions?

We can compare the inferred age of a radio source with the timescale
over which the environment is likely to change.

\begin{itemize}

\item The sound crossing time in gas of size $d$ is
$\approx 2 (d/{\rm kpc})~{(kT/{\rm keV})}^{-1/2}$ Myr.
This means that a medium $100~$kpc in size 
has not had time to change as a result of the presence of
a 100~Myr-old radio source, and a young radio source
should be in an environment which is similar to that of
its older counterparts.

\item The cooling time of gas of temperature $kT$ and density $n_e$ is
$\approx 3 \times 10^4$ ${(kT/{\rm keV})}^{1/2}~(n_e/{10^{-3}~\rm
cm^{-3}})^{-1}$ Myr.  Wide ranges of temperature and density relate to
a wide range of cooling times, in many cases approaching the Hubble
time.  If a cooling-flow is key to the fuelling of a radio source, as
suggested for powerful radio galaxies by \citeasnoun{Brem97}, then it
is curious that most radio sources appear to be 100~Myr old or
younger, where we might expect some to last for a Gyr or more.

\item The phases of development of an elliptical-galaxy atmosphere
\citeaffixed{Ciott91}{supernova wind, density increase, cooling, etc;
e.g.} are long compared with the measured
lifetimes of radio galaxies.  Are the host galaxies of
radio sources in one of these phases, or all?

\end{itemize}

\subsection{The Evidence}
\label{evidence}

\subsubsection{FRI radio galaxies}
\label{FRIs}

\begin{figure}
 \begin{center}
  \includegraphics*[width=14.0cm]{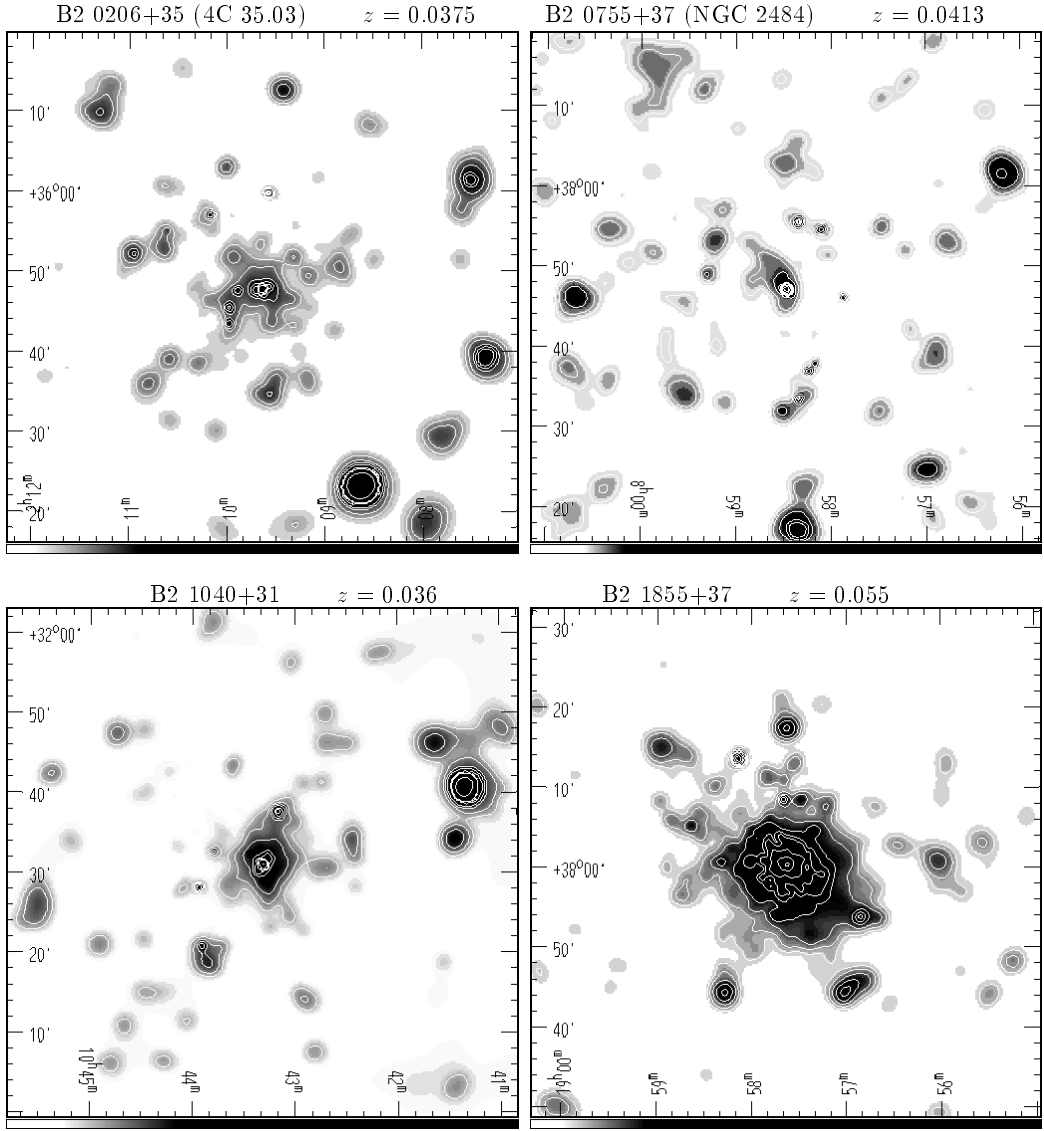}
 \end{center}
 \caption{Four representative $> 8$ ks-exposure ROSAT PSPC images
centered on B2 radio galaxies not in catalogued clusters.  Fields
are 1.3 square degrees, and 10 arcmin corresponds to between 600 and
900 kpc.  A range of sizes of X-ray-emitting atmosphere (group to
cluster dimension) is seen.  Figure adapted from Worrall \& Birkinshaw
(1999a).}
 \label{fig-b2xcont}
\end{figure}

\begin{figure}
 \begin{center}
  \includegraphics*[width=14.0cm]{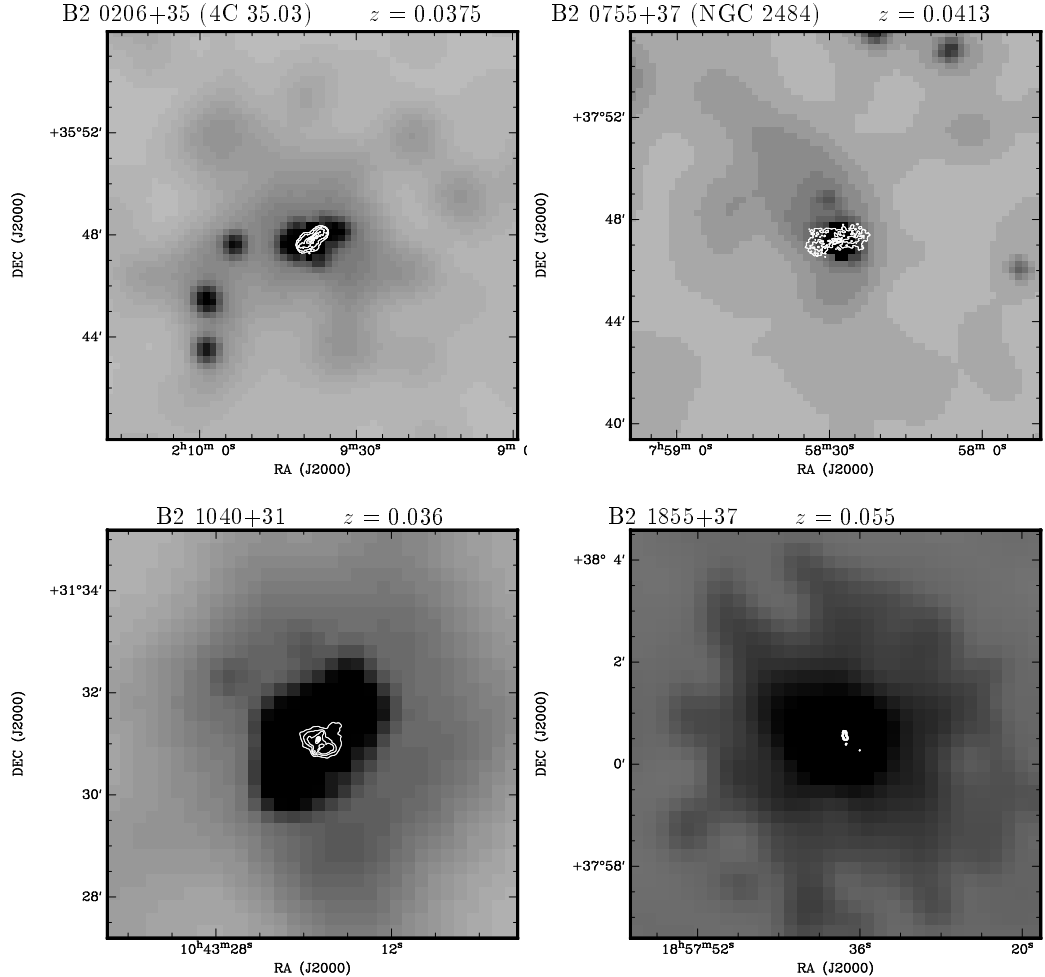}
 \end{center}
 \caption{The X-ray atmospheres (grey scale) of the B2-sample galaxies
in Fig.~\ref{fig-b2xcont} are substantially larger than
the VLA radio structures (contours).  Note the change of scale from
Fig.~\ref{fig-b2xcont}. Radio
data are at 5~GHz for NGC 2484 (image kindly provided by
M. Birkinshaw) and 1.4~GHz for the other sources (images kindly
provided by R.~Morganti).  }
 \label{fig-b2xr}
\end{figure}

Hot atmospheres have been detected around FRI radio galaxies with {\it
Einstein\/} and ROSAT.  For representative results I turn to the
largest sample of such objects with sensitive pointed X-ray
observations: the B2 radio-galaxy sample.  This is a 408~MHz
flux-limited sample of 50 radio sources identified with elliptical
galaxies of $m_{\rm Zw} \leq 15.4$~mag \cite{Coll75,Ulr89}, of which
40 were observed in ROSAT pointings, 39 being on-axis \cite{Canos99}.
Apart from one starburst galaxy and one BLRG, all are FRIs at $z \leq
0.072$.  Two of the galaxies are the dominant members of catalogued
Abell clusters (A1795 and A2199), and two are galaxies of the Coma
cluster.  The environments of other sample members are measured
through their X-ray observations, with particularly useful data for
sources observed with the PSPC \cite{Worrb99a}.
Fig.~\ref{fig-b2xcont} shows four representative X-ray images,
illustrating group to cluster scales typical of X-ray emitting
atmospheres.  Such gas engulfs the radio structures
(Fig.~\ref{fig-b2xr}), but the lack of correlation between
radio-source size, and size or central density of the X-ray-emitting
medium, means that the gas has indeed not had sufficient time to
adjust to the presence of the radio source (\S \ref{timescales}), and
it must be small-scale processes, on size scales less than those of
the overall gaseous environments, which are the major influence on
radio-source dynamics and propagation.

\begin{figure}
 \begin{center}
  \includegraphics*[width=8.0cm]{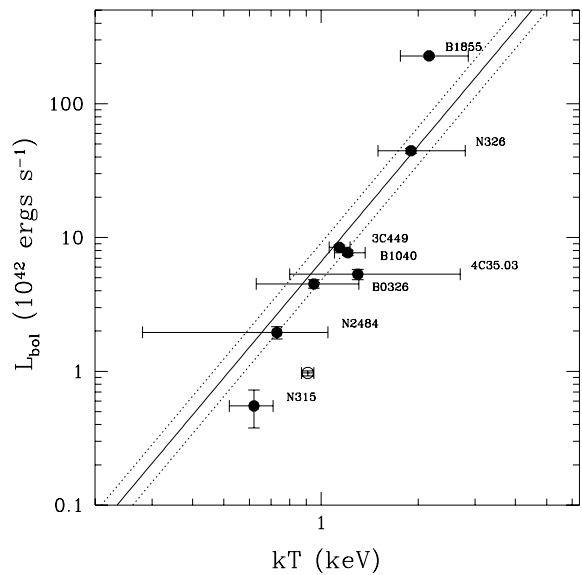}
 \end{center}
 \caption{The X-ray-emitting atmospheres of representative
B2 radio galaxies with
good ROSAT PSPC measurements fit an extrapolation of the
luminosity-temperature ($L_{\rm bol}$ -- $kT$) correlation for
more-luminous ($\sim 10^{44} - 10^{46}$ ergs s$^{-1}$)
optically-selected clusters (Arnaud \& Evrard 1999; dotted lines show
rms spread). This implies that the radio galaxy does not greatly
influence the gas fraction of the environment.  Figure from Worrall \&
Birkinshaw (1999a).}
 \label{fig-b2lt}
\end{figure}

Although the ROSAT PSPC's spectral resolution is poor by the standards
of the CCD detectors on ASCA, {\it Chandra}, and XMM, and of Astro-E's
calorimeters, the energy band is well matched to the typical
temperatures of groups and poor clusters.  The PSPC-derived
luminosities and temperatures of the environments of B2 radio galaxies
lie close to an extrapolation of the luminosity-temperature ($L_{\rm
bol}$ -- $kT$) correlation for more-luminous optically-selected
clusters (Fig.~\ref{fig-b2lt}).  Since $L_{\rm bol}$ is principally
governed by the gas mass, and $kT$ by the total gravitating mass, this
implies that the presence of the radio galaxy does not affect the gas
fraction of the environment.

The gas densities for the atmospheres of B2 radio galaxies do not
generally suggest the presence of cluster-scale cooling flows -- the
exceptions being for the two Abell clusters, A2199 and A1795.
A2199 is a particularly interesting case, where \citeasnoun{Oweil98}
have pointed out that the rotation measure of the core of B2~1626+39
(3C~338) implies appreciable central magnetic energy density,
complicating the interpretation of any cooling flow.
Possible galaxy-scale cooling flows, which may play a role in fuelling
the radio galaxies, need further investigation using
the sensitivity and spatial resolution now available with {\it
Chandra\/}.

\begin{figure}
 \begin{center} 
   \includegraphics*[width=12.0cm]{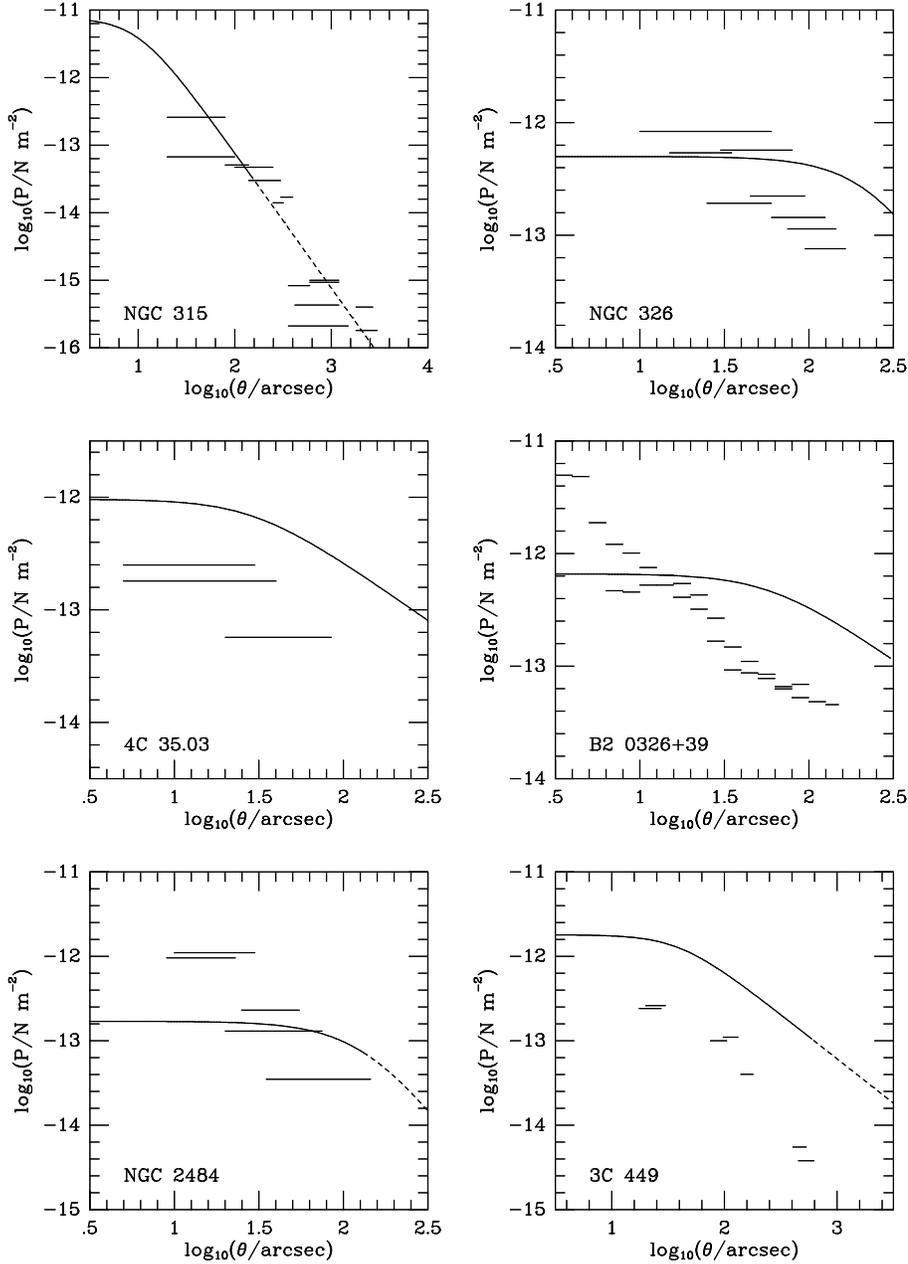}
  \end{center} 
  \caption{Thermal pressures in the atmospheres of six B2-sample radio
galaxies as deduced from fits to their ROSAT PSPC images (solid line,
shown dashed where extrapolated beyond region of clear X-ray
detection) compared with minimum internal pressure estimates
in the radio sources (horizontal bars). The intergalactic medium
is sufficient to confine
the outer parts of the radio structures, and in some cases even to
within 10~arcsec (5--10~kpc) of the core.  In the case of NGC~315 the
(extrapolated) pressure of the atmosphere matches the minimum pressure
in the radio source over a factor of $\sim 100$ in linear scale.
Figure from Worrall \& Birkinshaw (1999a).}
  \label{fig-b2press}
\end{figure}

\begin{figure}
 \begin{center}
 \begin{tabular}{cc}
  \includegraphics*[width=6.5cm]{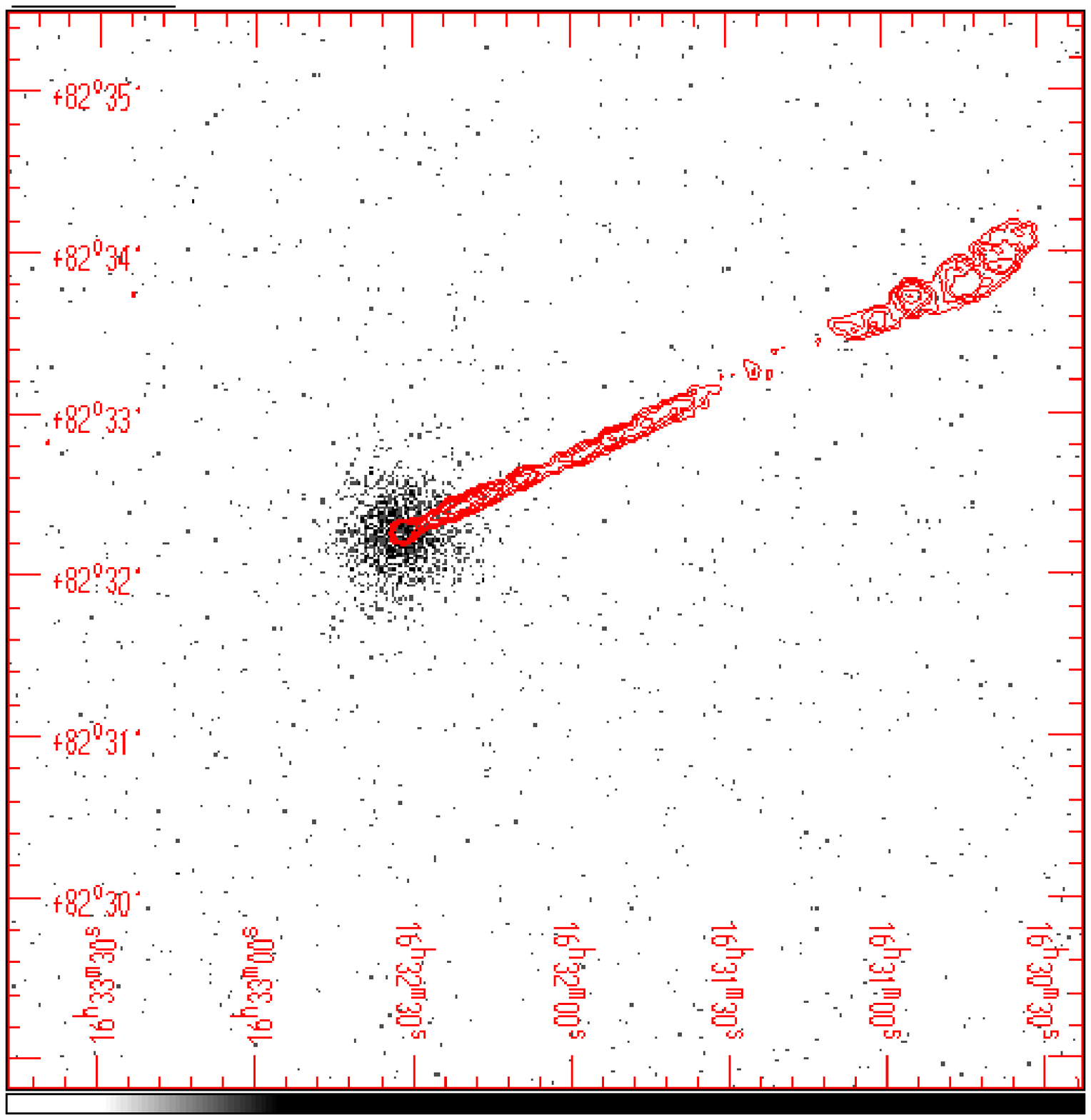}
  &
  \includegraphics*[width=6.5cm]{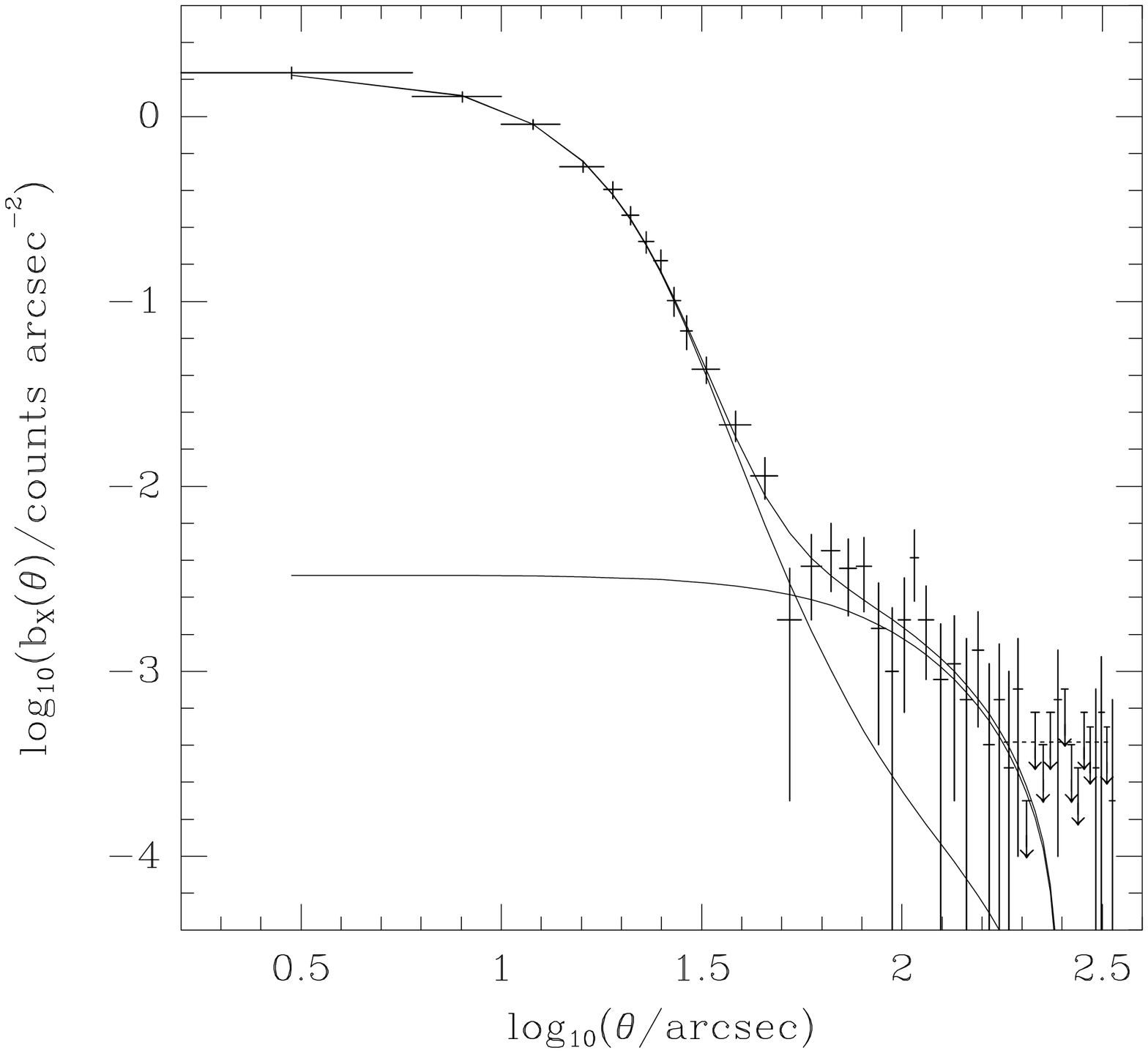}
  \\
 \end{tabular}
 \end{center}
 \caption{NGC~6251. 330~MHz radio contours on ROSAT PSPC image (left)
 and X-ray radial profile with best-fit model of unresolved emission
 plus weak group-scale gas described by a $\beta$-model (right).
 Radio jet features between 10~arcsec and 4.4~arcmin from the core are
 all overpressured with respect to the X-ray medium in this giant
 radio source.  Figure from Birkinshaw \& Worrall (1993).}
 \label{fig-6251x} 
  \vspace*{20pt}
\end{figure}

There is widespread evidence for pressure confinement of the kpc-scale
radio structures of FRI sources by the X-ray emitting medium
\citeaffixed{Morg88,Kill88,Fer95,Tru97}{e.g. Fig.~\ref{fig-b2press}
and}, and in some cases an apparent evacuation of the external medium
by the jets argues that additional internal jet pressure is required and
must be supplied by something other than thermal gas
\cite{Boh93,Hard-449-98}.  The exception to this picture, a
moderately low-power radio galaxy which appears to require an
alternative method of confining its long, straight, jet, is
NGC~6251 \citeaffixed{Birkw93,Wern99}{Fig.~\ref{fig-6251x} and}.

This review will not attempt a detailed discussion of how bending and
disruption of the kpc-scale jet structures of low-power radio galaxies
may relate to the motion of the radio galaxy through the gas or {\it
vice versa}.  However, various factors are likely to be influential,
including gas flows and density enhancements resulting from cluster
mergers \citeaffixed{Blit98}{e.g.}, density and temperature
discontinuities at the interface between the galaxy and cluster
atmospheres \citeaffixed{Sak99}{e.g.}, and buoyancy forces
\citeaffixed{Worr95}{e.g.}.

\subsubsection{High-redshift FRII radio galaxies}
\label{FRIIs}

\begin{figure}
 \begin{center}
  \includegraphics*[width=9.0cm]{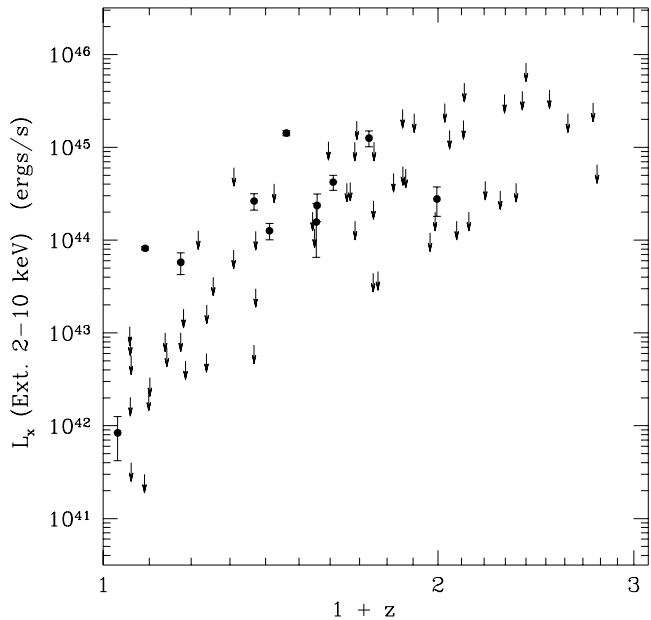}
 \end{center}
 \caption{The extended soft X-ray luminosity of high-power (FR~II)
quasars and galaxies from pointed observations of the 3CRR sample
(from the work of Hardcastle \& Worrall~1999a).
Detections, in order of increasing redshift, are galaxies 3C~98
and 3C~388, BLRG 3C~219, CSS quasar 3C~48, quasar 3C~215, galaxy 3C~295,
quasars 3C~334 and 3C~275.1, galaxy 3C~220.1, quasar 3C~254, and galaxy
3C~280. Upper limits are uncertain, particularly at low redshifts, due
to the need to model spatial extent and adopt a value for the gas
temperature.}
 \label{fig-frii}
\end{figure}

A major success of ROSAT has been the first detection of high-power
radio galaxies at high redshift.  Of the 38 radio galaxies at $z >
0.6$ in the 3CRR sample \cite{Lrl83}, 12 were observed in ROSAT
pointed observations and 9 were detected
\citeaffixed{Hardworr-3c99}{see summary in}, with the four most
significant detections exhibiting source extent
\cite{Worr94,Hard-220-98,Dick99}.  Moreover, extended emission is
detected around five 3CRR quasars at redshifts greater than $\sim
0.4$, one of which is at $z > 0.6$ \cite{Hardworr-3c99,Craw99}.
Fig~\ref{fig-frii} plots the extended luminosities for sources for
which the structure can be well modelled, together with upper limits
for the other 3CRR FRII sources observed in ROSAT pointings (roughly
half the sample). Powerful radio sources are finding some of the
highest-redshift X-ray clusters known to date, pointing to deep
gravitational potential wells early in the Universe.

\subsubsection{More local FRII radio galaxies}
\label{friis}

The nearer a source, the more likely it is that its various X-ray
emission components can be separated and the better will be the model
fitting to any extended emission.  FRII sources are rarer than FRIs
and thus typically more distant.  Cygnus~A and high-redshift FRIIs
with good X-ray data have extended X-ray luminosities one to two
orders of magnitude higher than a typical FRI, but what about other
more local FRIIs?  Their extended emission should be as easy to detect
if it really is so luminous.  The situation appears mixed, with the
extended luminosities for 3C~98 ($z = 0.0306$, $L_x \sim 10^{42}$ ergs
s$^{-1}$) and 3C~388 ($z = 0.0908$, $L_x \sim 10^{44}$ ergs s$^{-1}$)
differing by two orders of magnitude, and atmospheres for many sources
not yet detected (Fig~\ref{fig-frii}).  This luminosity range spanned
by 3C~98 and 3C~388 is similar to that of representative low-redshift
FRIs (see Fig~\ref{fig-b2lt}), although the full distribution of
extended X-ray luminosities for FRIIs is uncertain while many
nondetections remain.  Despite this, an interesting picture emerges.
Contrary to earlier work with less sensitive data \cite{Mill85}, the
X-ray atmospheres, where detected, provide sufficient pressure to
confine the radio lobes, with no disagreement from the many sources
for which only X-ray upper limits currently exist
\cite{Hardworr-envs99}.  In a detailed study of 3C~388,
\citeasnoun{LeaGiz99} have argued that that this implies the lobe
energy density is higher than given by minimum-energy arguments, and
they make the interesting point that if this is the case, jet
kinematic luminosities (normally calculated as energy density times
volume, divided by spectral age) are underestimated.

\subsubsection{Young Radio Galaxies}
\label{gps}

GHz Peaked Spectrum (GPS) radio sources are believed to be young FRII
sources and, even if only $\sim 100$~pc in size, the sound-crossing
time in the surrounding medium ($\sim 10^5$ years: \S
\ref{timescales}) is likely to be appreciable compared with the age of
the source \cite{Con20}.  We therefore expect the environments of such
sources to be similar to those in the inner parts of their older
counterparts. A search with ROSAT and ASCA for X-ray emission in or
around the archetypal GPS radio galaxy $2352 + 495$, at $z = 0.237$,
has set an upper limit for the soft X-ray band (0.2 - 2 keV) of about
$2 \times 10^{42}$ ergs s$^{-1}$ \cite{Odea96,Odea99}.  From
Fig~\ref{fig-frii}, this is already below the level at which the
atmospheres of some FRII radio galaxies are detected, suggesting that
slightly more sensitive observations with forthcoming missions should
see the atmosphere of this source.

\section{The Near Future}
\label{future}

A new era for X-ray astronomy has begun with the launch of {\it
Chandra\/} on July 23rd 1999, soon to be followed by XMM (December
1999) and Astro-E (January 2000).  Arcsecond imaging and detailed
spectroscopy with {\it Chandra\/} (see http://chandra.harvard.edu/)
will probe the central regions of radio galaxies, telling us whether
or not radio sources have dense cooling gas on sub-cluster/group
scales, a possible jet trigger, and will provide information on the
role of mergers and clumping in radio-galaxy formation, evolution, and
intermittency.  We expect new X-ray detections of knots, hotspots, and
compact jets, from which physical parameters of the emission regions
will be deduced.  XMM's unprecedented throughput coupled with $\sim
15$~arcsec imaging and CCD spectroscopy (see
http://astro.estec.esa.nl/XMM/) will measure large
samples of more distant sources, probing the relationship between
X-ray environment and radio-source structure. Spectroscopic separation
of components (non-thermal and thermal) will be easier, testing, among
other things, models for jet disruption.  The Astro-E mission (see
http://astroe.gsfc.nasa.gov/), with
its calorimeter spectral resolution of $\sim 12$~eV FWHM, should open
the door to new radio-source science -- kinematic and dynamical
studies via X-ray line emission -- paving the way for a future
generation of X-ray missions with high-throughput eV-level
spectroscopy (see http://constellation.gsfc.nasa.gov/ and 
http://astro.estec.esa.nl/SA-general/Projects/XEUS/).

\section{Summary}
\label{summary}

\begin{itemize}

\item Radio galaxies are multi-component X-ray emitters.

\item Jet-related emission dominates the soft X-radiation from the cores
of low-power radio galaxies, and may also be important in
high-power sources.

\item X-ray emission probes the magnetic field
strength in compact hot spots (via the synchrotron self-Compton process) and
extended regions (via inverse Compton scattering of CMB photons).

\item An extended X-ray medium is expected to
take a relatively long time to respond to the influence of a radio jet.

\item FRI radio sources reside in galaxy/group/cluster
hot atmospheres with densities which don't require cluster-scale
cooling flows, and pressures which are generally sufficient to confine the
radio jets.

\item High-redshift FRIIs may all reside in clusters with likely
cooling flows.

\item More local FRIIs may all be in environments which are rich
enough for radio-lobe pressure confinement.

\item There are many open issues which the new X-ray observatories
will imminently address.

\end{itemize}

\begin{ack}

I thank Mark Birkinshaw and Martin Hardcastle for their
continued collaborations and discussions, and their contributions to
this review.  I am grateful to the organizers of the workshop for
their invitation and partial support, and acknowledge further
support from NASA grant NAG5-1882.

\end{ack}

\end{document}